\documentclass{article}

\usepackage{titling}
	\newcommand*{\thetitle}{Research Challenges in Nextgen Service Orchestration}
\usepackage{amsfonts,amsmath,amssymb,amstext,latexsym,pifont}
\usepackage{adjustbox}
\usepackage{array,makecell}
\usepackage[inline]{enumitem}
	\setlist[enumerate]{leftmargin=3em,itemsep=0.33em}%
	\setlist[itemize]{leftmargin=3em,label={$\bullet$}}%
	\newlist{inlinelist}{enumerate*}{1}
	\setlist*[inlinelist,1]{%
		label=(\roman*),
	}
\usepackage{graphicx}
\usepackage{hyperref}
\usepackage{natbib}
	\setcitestyle{square,comma,numbers,sort&compress}
\usepackage[sc]{mathpazo} 
\usepackage[T1]{fontenc} 
\usepackage{microtype} 
\usepackage[hmarginratio=1:1,top=25mm,columnsep=20pt]{geometry} 
\usepackage[hang, small,labelfont=bf,up,textfont=it,up]{caption} 
\usepackage{booktabs} 
\usepackage{multirow}
\usepackage{float} 
\usepackage{abstract} 

\usepackage{titlesec} 
\renewcommand\thesubsection{\Roman{subsection}} 
\titleformat{\section}[block]{\large\scshape\centering}{\thesection.}{1em}{} 
\titleformat{\subsection}[block]{\large}{\thesubsection.}{1em}{} 

\usepackage{fancyhdr} 
\usepackage{color}
\usepackage{colortbl}
\usepackage[usenames,dvipsnames,table]{xcolor}
\usepackage{tikz}
\usepackage{xspace}
\usepackage{booktabs}

\makeatletter
\newcommand*{\etc}{%
    \@ifnextchar{.}%
        {etc}%
        {etc.\@\xspace}%
}
\newcommand*{\etal}{%
    \@ifnextchar{.}%
        {et al}%
        {et al.\@\xspace}%
}
\makeatother

\usepackage{authblk}

\author{Luis M.~Vaquero}
\affil{Faculty of Engineering, University of Bristol, UK}
\affil[ ]{\textit{Corresponding author:} \texttt{luis.vaquero@bristol.ac.uk}}

\author{Felix~Cuadrado}
\affil{EECS, Queen Mary University of London, UK}
	
\author{Yehia~Elkhatib}
\affil{MetaLab, SCC, Lancaster University, UK}
	
\author{Jorge~Bernal-Bernabe}
\affil{DIIC, University of Murcia, Spain}

\author{Satish N.~Srirama}
\affil{Institute of Computer Science, University of Tartu, Estonia}
	
\author{Mohamed~Faten~Zhani}
\affil{\'Ecole de Technologie Sup\'erieure, Montr\'eal}

\begin{document}

\date{}
\title{\thetitle}

\maketitle

\thispagestyle{fancy}

\begin{abstract}
    Fog/edge computing, function as a service, and programmable infrastructures, like software-defined networking or network function virtualisation, are becoming ubiquitously used in modern Information Technology infrastructures. These technologies change the characteristics and capabilities of the underlying computational substrate where services run (e.g. higher volatility, scarcer computational power, or programmability). As a consequence, the nature of the services that can be run on them changes too (smaller codebases, more fragmented state, etc.). These changes bring new requirements for service orchestrators, which need to evolve so as to support new scenarios where a close interaction between service and infrastructure becomes essential to deliver a seamless user experience. Here, we present the challenges brought forward by this new breed of technologies and where current orchestration techniques stand with regards to the new challenges. We also present a set of promising technologies that can help tame this brave new world.  
\end{abstract}


\section{Introduction}
\label{sec:introduction}
There is a new breed of technologies that are becoming mainstream in current Information Technology (IT) infrastructures. Fog computing aims to partially move services from core cloud data centres into the edge of the network~\cite{Vaquero2014}. Thus, edge devices are increasingly becoming an essential part of the IT infrastructure that extends from core cloud data centres to end user devices, allowing some management functions to be offloaded to the vicinity of sensors and other user devices, while heavy analytics can still happen in the cloud, possibly on aggregated data~\cite{Azimi2017}. This is especially relevant for resource-constrained churn-prone devices in the Internet-of-Things (IoT).

The fog has also been propelled by the advent of programmable infrastructures, like Software-Defined Networking (SDN), Network Function Virtualization (NFV)~\cite{Kreutz2015, GilHerrera2016}, and data centre disaggregation~\cite{Han2013, Klimovic2016, Rao2016}. These have simplified infrastructure configuration for data centre servers, storage, as well as core and edge networks. As a result, the infrastructure is able to adapt to the needs of the services that run on it, making the interplay between the services and~ the~infrastructure more dynamic and complex. 
 
In parallel, recent trends in software such as microservices, foster the utilisation of smaller software functions. Cloud-based serverless computing, also known as Function-as-a-Service (FaaS), is an attempt to tame complexity by dividing services into smaller individual functions that can be deployed and executed independently~\cite{Hendrickson2016,Baldini2017}. 

These technologies, shown in Table~\ref{table:conrevg}, gradually blend together to create a new IT environment characterised by the heterogeneity of equipment, technology and service, large-scale distributed infrastructures, high resource churn, and scarce computational power at the edge. Works that orchestrate serverless functions in an NFV context or amalgamating SDN and NFV orchestration are predominant in the first (left hand size) of the table. The central cell highlights efforts to blend programmable network techniques with fog orchestrators, while the right hand side cell shows works that try to make fog and serverless orchestration converge.

These technologies also affect the nature of services incurring smaller code bases and more fragmented state. The~complexity of the resulting IT environment and services makes service orchestration a central task to coordinate and schedule the operation of a myriad of distributed service components. Orchestration becomes even more challenging when different technologies are involved, requiring hybrid solutions that coordinate service provisioning and management taking into account the requirements and the particularities of each technology. While there has been some work~\cite{Liang2017,Sehgal2015,Suciu2015} on hybrid orchestration of pairs of these technologies, there has been no attempt to comprehensively tackle all of them. The orchestration challenges that result from this hybridisation of technologies are, therefore, still not fully understood.

\begin{table*}[]
\centering
\scriptsize
\caption{Summary of papers where these technology trends are converging}
\label{table:conrevg}
\begin{tabular}{ccc}
\toprule
\textbf{Programmable-FaaS} & \textbf{Edge/Fog-Programmable} & \textbf{Edge/Fog-Serverless}\\ \midrule
\cite{Roca2016,Jonathan2017,Abid2017} &
\cite{Cirani2015,Sehgal2015,Suciu2015,Lopez2015,Villari2016,Vilalta2016,Rotsos2016,Rostami2016, ohlen2016,Yigitoglu2017,Rehman2017,Truong2015,Consel2017,Liang2017,John2017,Mayer2017} & \cite{Martini2016,elkhatib2017microclouds,Dalla-Costa2017,Wen2017,Roca2017,Glikson2017} \\ \bottomrule
\end{tabular}
\end{table*}

In this work, we investigate research challenges in next-generation service orchestration frameworks. In particular, we present a comprehensive review with the~three following main goals:
\begin{enumerate}
    \item understand how each of the aforementioned technologies relying on orchestrators change requirements for orchestration
    \item reveal challenges for state-of-the-art techniques to meet those requirements
    \item discuss potential research directions to tackle new challenges in orchestration systems
\end{enumerate}

The rest of this paper is organised as follows. Section~\ref{sec:tech} introduces the main technologies where service orchestration is central to provision and instantiate services. Section~\ref{sec:reqs} analyses state-of-the-art orchestration techniques and introduces the main unsolved challenges. Section~\ref{sec:appro} introduces potential research avenues for these challenges. Finally, a critical discussion of the main lessons of this review work is presented in Section~\ref{sec:fin}.

\section{Recent Technological Trends}
\label{sec:tech}

\begin{figure}[!t]
\centering
\includegraphics[width=3.5in]{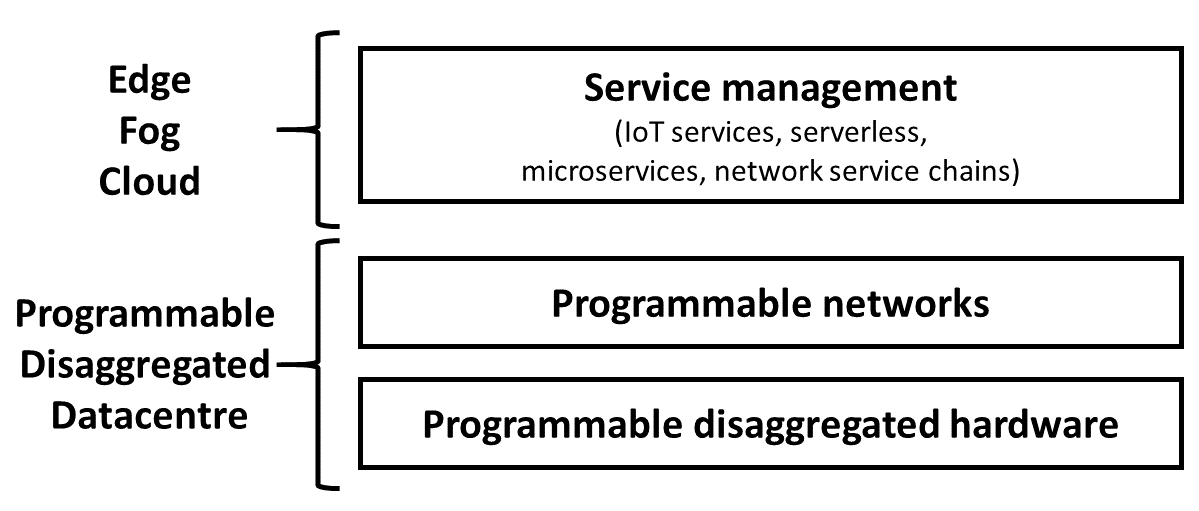}
\caption{Current State of the Orchestration Landscape. There is a much higher abundance of works dealing with orchestration as we go higher in this stack.}
\label{fig:layers}
\end{figure}

As can be observed in Figure~\ref{fig:layers}, most of the efforts around orchestration have happened around the cloud~\cite{Sousa2015patterns} and its natural expansion to the edge, via the fog~\cite{Jiang18}. Classic VM scheduling has partially been combined with edge resource selection and serverless functions in the data centre. As we will see, there are also some efforts on programmable network orchestration and very few dealing with the hardware, which new technologies make highly configurable and subject to orchestration too. In summary, orchestration is central at different levels from the data centre hardware and computing resources orchestration to the network management and service orchestration going from the core to the edge of the network.

\subsection{A Comprehensive Motivating Use Case on the New Technology Landscape}

A motivating use case of this technology hybridisation process is that of connected cars, which are estimated to produce between 4 and 100 TB of data a day~\cite{Xu2018}. Their potential need for upload bandwidth poses significant stress in current data and network infrastructures and edge/fog technologies have been postulated as a great starting point to cope with this huge data overload~\cite{Openfog,elkhatib2017microclouds}. 

These technologies can be complemented by smarter network management techniques, where an SDN controller may enable some traffic prioritisation for key data streams to nearby fog nodes (e.g. cars uploading updated information on road conditions to road side ``priority event'' relays). 

At night, benefiting from classic diurnal periodicity patterns in networks, Telco operators use WAN accelerators as NFV functions to speed up the offload of parked car data into car vendor-operated cloud services that may be running in a different continent. In this setting, car-vendor provided specific deduplication and encryption serverless functions can be used to minimise the amount of information to be uploaded in a secure way. In the same vein of data management, \cite{Rehman2017} proposed an architecture that blends edge, fog and IoT to provide analytic mechanisms for processing (and reducing/aggregating/synthesising) the amount of data that hits cloud servers. \cite{Cirani2015} proposed a fog node and IoT hub, distributed on the edge of multiple networks to enhance the implementation of several NFV services, like a border router, a cross-proxy, a cache, or a resource directory. 

Moreover, for cars driving in smart cities, it may be that CCTV cameras on lampposts may automatically detect blind spots for approaching cars and warn them about people or moving elements that can get into their trajectory. This is a mix of IoT sensors and SDN technologies. Similar safety features requiring local information and real-time processing have been achieved by combining an SDN controller with a Fog controller~\cite{Truong2015}. The fog offers network context information, location awareness, and ultra-low latency; which could satisfy the demands of future Vehicular Adhoc Network (VANETs) scenarios. 

All of these subsystems are critical to deliver a service where failure, delays, and security issues can be critical. Failing to synchronise the deployment edge nodes running serverless and NFV functions in a timely manner may result in critical security (e.g. unencrypted driver data, hacked cars), safety (malfunctioning SDN-enabled traffic prioritisation for critical events, like revealing moving objects in blind spots), functional (e.g. network clogging due to lack of deduplication or proper WAN acceleration resulting in outdated road maps) issues.

Each of these technologies \textit{per se} presents new challenges for orchestrators, but the combination of (programmable) infrastructure and services creates a novel interplay between infrastructure and service. 

In the remainder of this section, we analyse each of currently-deployed technologies separately to extract orchestration requirements. Our ultimate goal is to shed the light on the need for comprehensive orchestration techniques that could coordinate and schedule network services simultaneously through different technologies across the edge-to-cloud network.

\subsection{Programmable Disaggregated Infrastructures} 
\label{sub:pdi}

\subsubsection{Data Centre Hardware Disaggregation}
\begin{center}
	\textit{
	Classic single computer architectures have gradually been split apart, instead giving rise to disaggregated computers where CPU, memory, and storage are interconnected over a high-bandwidth network, rather than over a bus within a single chassis~\cite{Lim2009,Han2013}. 
}
\end{center}

Such physical separation allows for more flexibility in management and maintenance, catering for different and varying application needs~\cite{Barroso2009}; it also enables more efficient virtualisation of specific data centre resources~\cite{coglitore2014}. Under this model, data centres would not be composed of traditional servers each of them having its own resources (disk, memory, etc.), but they would instead consist of racks of specific resource types that are offered as pools of virtualised resources accessible via the network~\cite{Gao2016}. The granularity of these resources tends to be more finely controlled as there are strong incentives for cloud data centres to increase utilisation by encouraging use and release utilisation patterns, which require smaller execution units~\cite{Kilcioglu2017}. 
 
Several recent efforts follow this disaggregated model of data centre organisation, focusing specifically on flash~\cite{Klimovic2016} and traditional storage~\cite{Legtchenko2017}. While storage disaggregation is common practice (e.g. most cloud vendors offer some volume or elastic block store service), memory disaggregation is another novel trend~\cite{Rao2016, Gu2017, Caldwell2017}. Ideas on accessing remote memory were extensively studied 20 years ago and now they are getting revived due to the massive improvement in network latency (3 orders of magnitude). As memory latency has not improved as much, there is a gradual convergence of performance~\cite{Aguilera2017}.

Several hardware solutions are currently under development both as research projects~\cite{Lim2009,Lim2012,Syrivelis2017} and industry efforts, such as CCIX~\cite{CCIX}, Gen-Z~\cite{GENZ}, OpenCAPI~\cite{OpenCAPI}, and Omni-Path~\cite{OmniPath}. However, as data centres expand to the edge of the network, there will be a larger number of edge devices serving as infrastructure. This seems to indicate new protocols to access remote memory with limited (or none) hardware support are needed~\cite{Aguilera2017}. Edge devices and cloud will have to adapt to support more abstract configuration and programmability mechanisms.

\subsubsection{Programmable Memory and Storage}
Disaggregated service infrastructures aim at increasing the utilisation of available resources by separating resources in different pools. This is common in current data centres. 
\begin{center}
	\textit{
	These disaggregated elements are also becoming configurable by software. For instance, storage volumes are dynamically attached to the memory/cpu of a VM when users request it. 
}
\end{center}
 
Storage systems have improved in speed significantly in recent years~\cite{Gray2006}. This is largely thanks to low-latency storage (e.g. flash) that moves the bottleneck to be the CPU and network instead of storage where it has traditionally been. Additionally, the rise of \textit{de facto} APIs for network-attached storage systems, such as OpenKinetic key-value stores\footnote{https://www.openkinetic.org/}, further contribute to this recent development.
 
Recent work has focused on developing programmable storage systems that allow the composition of new storage services through the reuse of existing storage interfaces~\cite{Sevilla2017}. This is different from the notion of Software Defined Storage (SDS)~\cite{Stefanovici2017} where storage racks are assembled from commodity hardware. 

Memory disaggregation together with the advent of non-volatile memory (NVM) technologies and optical interconnects~\cite{Faraboschi2015} are also requiring higher memory programmability -- see  Table~\ref{table:needs}. In some systems, like HPE's `The Machine'\footnote{https://github.com/FabricAttachedMemory/tm-librarian} or FluidMem~\cite{Caldwell2017}, compute resources can access additional memory in a data centre on demand. In general, the ``softwareisation'' of the infrastructures enables multi-tenant usage, which will have to be considered by orchestrators when planning resource allocation (see `multi-vendor / -domain' in Table~\ref{table:needs}). The possibility of accessing byte addressable memory in nearby edge devices over wireless communications seems less feasible due to excessive latency in current technologies.

As indicated in Figure~\ref{fig:layers}, the number of works dealing with systematic orchestration of disaggregated and programmable hardware is very limited, mainly due to their recent appearance and limited industrial adoption at scale.  

Programmable storage systems allow encapsulating storage functionality as reusable building blocks, leveraging storage capabilities through new interfaces. However, the composition of interfaces is complex, and it is important to limit the usage of object interfaces, as is the sandboxing of the runtime space~\cite{Sevilla2017}.

\subsubsection{Programmable Networks}
A few years ago, network service management used to be highly manual, resulting in extensive capital and operational costs.
A good illustration showing how daunting is to manage network services is the example of provisioning network service chains in traditional networks.
A network service chain is an ordered series of network functions (e.g., routers, Firewalls, intrusion detection systems) that process incoming  traffic. Traditionally, provisioning such a chain requires a lot of endeavour from IT operators to acquire and deploy networking equipment and to manually configure the network to steer the incoming traffic across the service chain components.
These tasks not only require weeks to months to be implemented by also need knowledgeable human resources and a lot of effort. 
Furthermore, the need for new hardware equipment and human resources incur high capital and operational costs, making this simple service provisioning operation extremely daunting, costly and time-consuming.

The advent of SDN and NFV technologies brought provisioning time from the scale of weeks and months to that of minutes and significantly reduced human intervention by automating all the service chain provisioning steps. On one hand, SDN has succeeded to revolutionise the way network components are configured and managed. In traditional networks, each networking equipment consists of:
\begin{inlinelist}
    \item a control plane, that is mainly responsible for taking routing decisions; and 
    \item a data plane, that is in charge of forwarding traffic according to the decisions made by the control plane.
\end{inlinelist}
SDN technology moves the control plane from the network equipment to a logically-centralised software-based controller, making it possible to manage the whole network from a single point of control. This offers network operators the programmability and the flexibility, allowing to easily configure their networks and dynamically adapt their routing paths to applications' performance requirements. In addition, SDN provides the tools to set up bespoke control over the traffic, allowing to define fine-grained flow forwarding rules (e.g., flow definitions are easily performed by the network operator using diverse information in packet header fields). 

On the other hand, NFV is a new technology that leverages server virtualisation technology to turn network functions (e.g., routers, firewalls, proxies) that traditionally used dedicated hardware (middleboxes or network appliances) into software that run on top of general purpose hardware such as virtual machines (VMs)~\cite{GilHerrera2016}. NFV inherits all the advantages of virtualisation \cite{ZHANISurvey2013}. For instance, it offers the possibility to create different types of network functions whenever needed and to adjust their processing capacity to the varying demand. It also allows to easily create a copy of a network function or simply migrate it to another location if and when required.

The combination of SDN and NFV technologies makes it now possible to provision within very short timescales a fully-fledged network service chain. In~this context, service chain provisioning and orchestration is one of the most challenging problem as it requires solving several optimisation problem simultaneously (e.g., \cite{Racheg-ICC2017}).
Challenges include placing virtual machines, connecting them and steering the traffic through the ordered chain of the network functions. This is a non-trivial task as it requires finding the best compromise between VMs' hosting costs,  bandwidth costs, and performance requirements in terms of the chain end-to-end delays~\cite{Huin2017,Ghrada-PVESDN2018}. 

The orchestration framework needs also to manage other features like multi-tenancy, fault-tolerance and availability management. Multi-tenancy allows multiple users to share the same infrastructure and hence requires resource isolation between different service chains and better performance management to satisfy each tenant's requirements. The orchestrator needs also to ensure high service chain availability through efficient fault-management (e.g., \cite{SurvBook2015,RabbaniIEICE13,ZhangVenice2014}). This requires leveraging SDN and NFV technologies to put forward a set of solutions allowing to handle different types of failures (e.g., node, link and software failures) and mitigate their impact on service chain availability.

\subsection{Edge and Fog Computing}
\label{sub:edge_fog}

\begin{center}
	\textit{
Edge and fog technologies shift centralised cloud computations toward edge devices in order to decrease latency, improve spectral efficiency, enable enhanced context-specific functionality, and support the massive machine-to-machine type of communication. They also allow for localised functions such as processing that benefit from p2p-style communications and exploiting nearby resources~\cite{Ciobanu2017}, exploiting data locality to obtain faster results.
}
\end{center}
This is driven by the explosion in the number of connected devices and services, and the increasing demands of applications for low latency and interactive experiences~\cite{Wen2017,Jiang18}. 
 
Additionally, fog computing is facilitated by the availability of suitable hardware in the form of small, affordable, low-power computers~\cite{elkhatib2017microclouds}, along with improvements in virtualisation technologies that enable slicing resources between different users to provide isolated environments. This contributes to a need to support highly heterogeneous hardware and software stacks that emanates from fog/edge environments and is improved thanks to the usage of programmable infrastructures (Table~\ref{table:needs}).
 
There is a need for a close interplay between cloud and network resources, where resources can be dynamically instantiated. For example, \cite{SaeLor2012} describes how to dynamically deploy cloud storage services in core networks to simplify data backups across data centres. This example requires: 
\begin{inlinelist}
    \item Partial view of the capabilities of other ISPs (ability to select services and providers);
    \item Dynamically instantiate/select virtual resources (e.g. virtual routers and storage VMs);
    \item Ability to guarantee predetermined QoS levels across all the services composed to deliver a network function~\cite{Martini2016};
    \item Monitor quality metrics and automatically re-configure the VNF, if needed.
\end{inlinelist}
 
Similar techniques are being deployed at the edge. The main cloud vendors (such as Amazon and Microsoft) are making it easy to integrate IoT devices into their clouds mainly by providing gateway management and tight integration with their other cloud services~\cite{Greengrass2017, Azure2017, Kura2017, Liota2017}.

While there are interesting works coping with resource provisioning, deployment, and scaling of applications on edge devices~\cite{Wang2017,liyanage2016mepaas}, there is still a gap that needs to be filled before accomplishing the vision of a more decentralised infrastructure where the devices can synchronise themselves without the need of central cloud services~\cite{Vaquero2014}. This is where the usage of serverless functions may come in handy, since they can be run at the edge on locally generated data,  and rely on the cloud for management, analytics, and durable storage~\cite{Greengrass2017}. The orchestration of these small functions that run scattered across the edge of the network and the interaction with centralised cloud services is still a challenge~\cite{Wen2017} (larger-scale and finer-grain required -- see Table~\ref{table:needs}). 
 
While serverless frameworks  provide facilities to define dataflow orchestration~\cite{giang2015developing,Ravindra2017}, they generally ignore how to enable inter device communication. For instance, a set of devices may choose another, more powerful, neighbouring machine to do some aggregation before sending the data to the cloud. Also, current orchestration approaches do not cope well with device churn (another essential requirement for orchestrators -- see Table~\ref{table:needs}), shadow devices being one of the few solutions out there~\cite{Greengrass2017}. 
 
\cite{Skarlat2016} defined fog cells, as single IoT devices coordinating a group of other IoT devices and providing virtualised resources. These resources are located close to the edge of the network next to the data sources or sinks, instead of involving the cloud. They build on the idea of fog colonies (also referred to as edge clouds by~\cite{Vaquero2014}), which are micro data centres made up from an arbitrary number of fog cells. Fog colonies distribute task requests and data between individual cells, allowing for cloud offloading and multi-cloud deployment~\cite{Elkhatib2016crosscloudmap}. 

We envision colonies and cells as dynamic entities that can be formed based on short term convenience: for instance, summer campers establishing a mesh network with nearby devices to enable local connectivity in areas with poor service (e.g. each phone relays nearby messages)~\cite{Wen2017}. We refer to these dynamic and transient cells and colonies as fluid distributed organisations. A fog node can start autonomously, and become the lead orchestrator or part of a set of distributed orchestrators reaching some quorum before making decisions and dynamically leave the colony without warning~\cite{Brito2017}.

In this setting, hierarchical approaches tend to fall quite short; their scalability and manageability become harder as scale, distribution, and churn increase~\cite{Hong2013}. Also, most prior approaches tend to assume that a device participates in a single cell or colony, while in reality it may belong to several of them at the same time. Moreover, multi-tenancy is a very uncharted land (see Table~\ref{table:needs}). As edge devices access the network via different ISPs and potentially operate across various administrative domains~\cite{elkhatib2015building}, the ability to deal with multi-organisation/multi-tenant environments becomes a must for any orchestration technology.
 
Edge/fog devices run on energy-efficient low spec'ed hardware running smaller execution units. The basic units of execution have continuously shrank: from VMs to containers, microkernels, and serverless functions. 
Such smaller execution units are dictated by the need for increased flexibility and control, and also the strong economic incentives to increase resource utilisation under usage variability~\cite{Kilcioglu2017}. As a consequence, the number of lines of code associated with each unit of execution, the duration of each execution unit, and the amount of state it keeps have all greatly decreased. More devices (larger scale) come together with finer grained units of execution (data and code) (see Table~\ref{table:needs}).

Radio access, network transport, and cloud resources are coordinated by a higher level orchestration layer~\cite{Rotsos2016, Rostami2016, ohlen2016, John2017}. These three different resource types are a simplification of a much more complex mixture of networking technologies and protocols~\cite{Guerzoni2017}. Fog and edge computing create a scenario of unbounded heterogeneity, or \textit{hyper-heterogeneity} as it affects devices, software stack, communications, and data management, and hardware technologies.

Orchestrating the security in Edge-Fog computing and IoT is a huge challenge. It encompasses different  technologies from different fields, including, among others, wireless-cellular networks, distributed systems, virtualisation, and platform management \cite{Satyanarayanan}. It imposes new models of interaction among different heterogeneous clouds, which require mobility handover and migration of services at both, local and global scale. The threats from all those building blocks are inherited in the edge-fog paradigm. Different layers are affected by diverse technologies that need to interoperate to achieve holistic and confidential connectivity \cite{ROMAN-iot}.

At the edge, services and devices can be compromised through different attacks such as privilege escalation, service manipulation or even physical damages in unprotected edge/fog/IoT environments. At the network level, edge-fog and IoT impose additional threats, as wireless connections might provoke man-in-the middle attacks~\cite{stojmenovic2014fog}, spoofing attacks, eavesdropping or traffic injection, or authentication~\cite{Osanaiye2017} to name a few. Privacy leakage is also an important threat at the edge, as mobile devices can be tracked without user awareness. In this sense, Roman et al.~\cite{ROMAN2018680} have recently identified main security threats and challenges in mobile edge and fog computing.

Aspects such as certification of virtualised applications, tenants data isolation and sharing, resource usage control still require the definition of edge device policies and specific access control mechanisms~\cite{Haus2017,Bernabe2017}. It is hard to establish a chain of trust with edge devices under different organisational boundaries. Dynamic and multidimensional trust and reputation access controls mechanisms have been suggested~\cite{taciot}. Also, privacy-preserving mechanisms, satisfying ''unlinkability'' and minimal disclosure properties \cite{Bernabe2017}, as well as data aggregation schemes to conceal user data, are needed in computing-enhanced IoT applications \cite{lu2012eppa}. 

Expressive languages for defining high level security policies and models could serve as input for orchestrators to organise and choreograph the aforementioned security services.

\begin{table*}[]
\scriptsize
\centering
\caption{Emerging Orchestration Needs in New Technologies}
\label{table:needs}
\begin{tabular}{cll}
\toprule
\textbf{Technology} & \multicolumn{1}{c}{\textbf{Functional Orchestration Needs}} & \multicolumn{1}{c}{\textbf{Requirement}} \\ \midrule
\multirow{9}{*}{\textit{NFV}} & quick routing adaptation~\cite{Rotsos2016, Mijumbi2016, Katsalis2016} & \thead{\scriptsize Dynamism and Speed} \\ \cline{2-3} 
                              & heterogeneity in infrastructure and VNFs~\cite{Rotsos2016}  &  \thead{\scriptsize Heterogeneity} \\ \cline{2-3} 
                              & reduced human involvement~\cite{Rotsos2016}  &   \thead{\scriptsize Automation}  \\ \cline{2-3} 
                              & comprehensive NF, lifecycle management; virtual tenants~\cite{Mijumbi2016,Mijumbi2015,Munoz2015} & \thead{\scriptsize Multi-tenant} \\ \cline{2-3} 
                              &  coordination across orchestrators~\cite{Katsalis2016,Mijumbi2015}, move to edge~\cite{Katsalis2016} & \thead{\scriptsize Multi-domain} \\ \cline{2-3} 
                              &  single function scheduling~\cite{Mijumbi2015,Dalla-Costa2017} & \thead{\scriptsize Finer grain}\\ \cline{2-3}
                              &  function splitting~\cite{Mijumbi2015,Dalla-Costa2017}, deployment \& config checks~\cite{ohlen2016} & \thead{\scriptsize Larger scale}\\ \cline{2-3} 
                              &  Virtualisation issues \cite{vaughan}, security scalability, \cite{Rotsos2017} &\\ 
                              & availability  \cite{etsi-nfv-security}, vulnerable applications ~\cite{Scott-Hayward2013}, topology validation \cite{etsi-nfv-security} & \thead{\scriptsize Security}\\ 
                              \midrule
\multirow{4}{*}{\textit{SDN}} & controller planning~\cite{Banikazemi2013,Mijumbi2015,Munoz2015}, move to edge~\cite{Vilalta2016} &  \thead{\scriptsize Multi-domain} \\ \cline{2-3} 
                              & heterogeneous network~\cite{Munoz2015}, end-to-end connectivity~\cite{Guerzoni2014}, move to edge~\cite{Vilalta2016} & \thead{\scriptsize Heterogeneity} \\ \cline{2-3} 
                              &  end-to-end connectivity~\cite{Guerzoni2014}, move to edge~\cite{Vilalta2016}, global path computation~\cite{Lopez2015}  & \thead{\scriptsize Larger scale}  \\ \cline{2-3}
                              &  east-west confidentiality \cite{etsi-nfv-sdn}, availability \cite{Shin:2013}  & \thead{\scriptsize Security}  \\ \cline{2-3} 
                              &  blend SDN/NFV orchestration~\cite{Mijumbi2015,Munoz2015}  &  \thead{\scriptsize Multi-technology} \\ \midrule      
\multirow{2}{*}{\thead{\textit{\scriptsize Programmable} \\ \textit{\scriptsize storage/memory}}} & composition of storage service though separate storage modules~\cite{Sevilla2017} & \thead{\scriptsize Scalability, Finer-grain}  \\ \cline{2-3} 
                              & remote/disaggregated (edge) memory~\cite{Faraboschi2015,Caldwell2017,Aguilera2017} & \thead{\scriptsize Multi-vendor, Multi-domain} \\ \cline{2-3} 
                              
                               & protection of new programmable storage interfaces, sandboxing~\cite{Sevilla2017} & \thead{\scriptsize Security}
                              
                              \\ \midrule

\multirow{6}{*}{\thead{\textit{\scriptsize Fog/Edge} \\ \textit{\scriptsize computing}}} & dynamic coalitions of edge devices and cloudlets ~\cite{ Skarlat2016,Vaquero2014,Brito2017,Wen2017,Pradhan2017,Amjad2017,Ciobanu2017,chang2017mobile,Jiang18} & \thead{\scriptsize Dynamism, Churn, \\ \scriptsize Scalability, Locality-awareness}  \\ \cline{2-3} 
                              & going beyond shadow devices for reliability~\cite{Greengrass2017,Brito2017,Wen2017,Pradhan2017}  & \thead{\scriptsize Churn } \\ \cline{2-3} 
                              & dynamic end-to-end service availability ~\cite{SaeLor2012,Brito2017}  & \thead{\scriptsize Multi-tenant, Multi-domain}  \\ \cline{2-3} 
                              & smaller execution units~\cite{Kilcioglu2017,Wen2017,Amjad2017,liyanage2016mepaas}  & \thead{\scriptsize Larger scale, Finer grain}  \\ \cline{2-3} 
                              & diversity~\cite{Rostami2016, ohlen2016, John2017,Guerzoni2017}  & \thead{\scriptsize Heterogeneity} \\ \cline{2-3} 
                              &  M2M confidentiality, wireless-based attacks \cite{ROMAN2018680}, trust management \cite{taciot} & \thead{\scriptsize Security} \\
                              & AAA ~\cite{Osanaiye2017} \cite{stojmenovic2014fog},  privacy-leakage \cite{lu2012eppa}.  & \thead{\scriptsize privacy} \\ \cline{2-3} 
                              & ensure quality-of-service on a variety of infrastructure elements~\cite{Martini2016,Jiang18}  & \thead{\scriptsize Heterogeneity, Multi-domain}  \\ \midrule
\multirow{3}{*}{\thead{\textit{\scriptsize Serverless} \\ \textit{\scriptsize computing}}} & reduce latency in function execution and state handling~\cite{Jonas2017,Baldini2017} & \thead{\scriptsize Speed, Locality-awareness}  \\ \cline{2-3} 
                              & going beyond shadow devices for reliability~\cite{Greengrass2017}  & \thead{\scriptsize Churn}  \\ \cline{2-3} 
                              & FaaS and Serverless security issues \cite{pdjarny}  & \thead{\scriptsize Security} \\ \cline{2-3} 
                              & smaller execution units, smaller state~\cite{Hendrickson2016,StepFunctions2017}  & \thead{\scriptsize  Larger-scale, Finer-grain}  \\ \bottomrule
\end{tabular}
\end{table*}

\subsection{Serverless Computing}
\label{sub:faas}

For over a decade, large-scale complex computations have shifted to a high-level, function-oriented model in which computation is expressed as functions that are composed into dataflows, and automatically deployed and managed by a cluster~\cite{Dean2004,Zaharia2010}. From the point of view of a developer, this level of abstraction splits the jobs that are to be executed from the way they are provisioned. Resources are automatically freed when computing jobs are done.

\begin{center}
	\textit{
Similarly, the concept of ``serverless computing'' (FaaS) refers to server-side logic written by the application developer, running in fully managed stateless compute containers that are event-triggered, and ephemeral (may only last for one invocation).
}
\end{center}

Fundamentally, the serverless paradigm completely decouples running code from the management of the supporting server applications. This is a key difference compared to other modern architectural trends like containers and Platform as a Service (PaaS). PaaS/Containers applications are not geared towards bringing entire applications up and down for every request, whereas FaaS platforms do exactly this.
 
From the provider's perspective, many cloud deployments tend to be very static as users deploy VMs for long periods of time. In contrast, the variation in memory and CPU utilisation tends to be much more variable~\cite{Kilcioglu2017}, and thus offers a better way to optimise resource usage and to bill users. Therefore, cloud vendors have strong incentives for services to be built on serverless architectures as opposed to following fixed-price model for long-running VMs~\cite{Hendrickson2016}. The billing model is based on the number of function invocations and how many GB-s the function uses. Thus, developers have strong incentives to build smaller functions that minimise execution time and memory consumption. This is a factor contributing to the `finer-grain/large-scale' requirement in Table~\ref{table:needs}.
 
Function initialisation and state recovery/storage are key elements that may slow down function execution, resulting in potentially more expensive executions or totally failed function chains~\cite{Jonas2017}. These contribute to the ``Speed'' and ``locality awareness'' requirements of next generation orchestrators (see Table~\ref{table:needs}).
  
Delivering cloud applications typically means composing different tasks (hence the need to orchestrate function composition that comes as a requirement in Table~\ref{table:needs}). One way to deliver cloud applications in a serverless way would be creating a function for each task. However, orchestrating those small functions (finer granularity than a microservice) could be really hard to debug and optimise~\cite{StepFunctions2017}.
 
Initialisation latency, state push/pull, and availability problems become significantly worse as we move to an edge/fog computing arena~\cite{Hendrickson2016,Jonas2017,Baldini2017}. Being able to control co-location of fine-grained code with tiny subsets of the data may prove essential to deliver appropriate orchestration capabilities in serverless environments  (see Table~\ref{table:needs}). The ubiquity of edge devices and the advent of fog computing, together with the finer-grained nature of typical serverless functions, have changed the options available for orchestrators to play a key role.

FaaS moves some of the security concerns from the user to the platform provider. As it is remarked in \cite{pdjarny}, users do not need to take care of OS patches anymore, but security updates to 3rd party dependencies of applications remains the same. In FaaS servers are immutable and short lived, minimising the possibility of a long lived compromised server. However, security monitoring and accounting and debugging becomes more difficult \cite{Baldini2017}, as traditional monitoring agents need to be exposed by FaaS providers.

\subsection{Current Standardisation Efforts}
\label{sub:standard}

The term orchestration is used pervasively in the literature reviewed so far. The end result is a myriad of often incompatible standards that tend to cover one of the new technological trends. In this subsection we present the most prominent approaches related to these trends.

The ETSI NFV Management and Orchestration (MANO) is one of the most solid pieces of work in terms of NFV standardisation~\cite{mano}. \cite{Brito2017} have suggested that NFV should be the starting point for new standardisation efforts beyond NFV, given the  need  to  define, for each application domain, the  scope, properties, and  requirements of service orchestration concepts is not  different  in  the  Fog  Computing  environment. This suggestion has also been followed by project Tacker in an attempt to integrate a MANO-compliant orchestrator on top of one of the most widely-used cloud management suites, OpenStack\footnote{https://wiki.openstack.org/wiki/Tacker}. ETSI MANO deals with computing nodes where only CPU, memory, network, hypervisor, and Operating System can be chosen. In the edge/fog, devices are much more heterogeneous and have capabilities other than that(e.g. sensors and actuators). These unreliable devices play a major role in the process and need to be taken into consideration in the architecture or an orchestrator.

The ETSI Mobile Edge Computing (MEC) Reference Architecture~\cite{emc} emphasises the need to consider a comprehensive set of constraints, and also refers to triggered application instantiation and relocations, one of the characteristics of our solution, similar to a fog orchestrator~\cite{Brito2017}.  However, a detailed specification or/and Reference Architecture for the orchestrator is still missing.

The Open Networking Foundation (ONF) has been working on SDN standardisation for quite some time. Most of their documents treat orchestration as an external (client-driven) coordinator of several SDN controllers~\cite{ONF}. This is similar to the recently released OpenFog Reference Architecture~\cite{Openfog}, which mainly focuses on the fog node.  

The Topology and Orchestration Specification for Cloud Applications (TOSCA) is turning into the \textit{de facto} standard for modelling service orchestration~\cite{TOSCA}. TOSCA is especially suited for defining services, their building blocks, requirements and capabilities, but it still does not help solve problems like device churn, multi-domain/multi-tenanted orchestration of tiny functions at scale. In addition to TOSCA, there are a few others like the IETF NETCONF Data Modelling Language (NETMOD) WG, together with recent expansions for VNF network services (following the ETSI MANO architecture). More acronyms would be needed for a top-down end-to-end solution (e.g. cloud VM configuration languages, network traffic engineering configurations, etc.).

Current trends for service description and modelling are very prescriptive and fragmented. Recent standardisation efforts for the new technologies presented in this section are poorly specified and cannot cope with all the new requirements.

\section{Current Orchestration Challenges}
\label{sec:reqs}

Building on the previous section, here we analyse the requirements in Table~\ref{table:needs} in depth, trying to systematically unveil orchestration challenges and attempts to tackle them.  

\subsection{Churn and Unreliability}
Edge resources are inherently  volatile. The advent of the fog extends the cloud to the edge to a point where end user devices could be employed as infrastructure to deploy services or service functions~\cite{Ciobanu2017}. Moreover, as the fog is increasingly being used to support transient FaaS, e.g. to support context-specific mobility functions, functions are ephemeral which imposes a rate of change much higher than in cloud environments. 

Such nature poses significant challenges on different functions that enable orchestration. Description of resources and functionality might not always be accurate as it could quickly be outdated, which complicates reliable deployment and service level agreement (SLA) guarantees. Similarly, discovery must be dynamic in order to take advantage of new resources as they become available, as well as move away from decommissioned/failed resources (see Table~\ref{table:acc}). Further, monitoring needs to always seek up-to-date information, avoiding obsolete data and empirically complementing self declaration from devices. 

Discovering mobile edge devices by scanning all connected communication interfaces and enlisting all locally available mobile edge devices is a first common approach~\cite{Rehman2017}. An example of this prevalent approach is Foggy~\cite{Yigitoglu2017}, that relies on a centralised orchestration server and a container registry for deployment. However, high churn may advise against this practice since it may exhaust remote device batteries or increase the energy bill of the infrastructure provider(s).

In contrast to the highly stable resource provisioning in cloud data centres, edge devices can be switched off dynamically, in order to cater to edge workloads and latency-, location- and privacy-specific requirements. As such, the edge is a highly volatile operational environment where resource availability is liable to significant changes over time and is divided across multiple domains of those operating the edge resources. The fog acts as a stabilisation layer, offering more reliable infrastructure in the proximity of the edge devices. Still, dealing with orchestration in unreliable environments comes with specific challenges to each of the phases of the orchestration process.

\subsection{Heterogeneity}

Inherent in the task of orchestration is the dealing with resources of varying nature and access methods, and that are managed under different administrative domains. Furthermore, the fog paradigm offers an alternative to the centralised model of the cloud. As such, any attempt to resolve the above challenges through central elements to handle monitoring, scheduling, configuration, etc. would undermine the benefits of disaggregation~\cite{giang2015developing,Brito2017}. 

These challenges call for two main approaches to orchestration. First, distributed orchestration is essential to deliver the potential of the fog paradigm, where orchestration elements manage different edge domains and inter-coordinate in a hierarchical or peer-to-peer fashion. Currently available tools, such as IBM Node-RED\footnote{\url{https://nodered.org/}}, offer high-level developer tools for creating interconnected flows. However, they are tailored specifically towards IoT functions. More generic tools are needed in order to support rich and customised coordination between a distributed network of orchestrators. Second, a sophisticated level of abstraction is needed to hide away the complexity of heterogeneity from application development and deployment processes. Toolsets are needed to not just simplify the tasks of resource discovery and monitoring, but also end-to-end lifecycle management, and to compose elaborate adaptive migration policies and mechanisms.

Managing heterogeneous resources across distinct administrative domains is already a challenge, but the independence between resource management and workload scheduling on fog devices increases the difficulty of their orchestration.

\subsection{Dynamism}
The IoT brings \textit{ad-hoc} devices at the edge of the network as infrastructure to run services. The quickly changing network conditions at the edge bring a significant amount of additional dynamism to service-based applications, in contrast with the relative stability of large data centres. Dynamic adaptation mechanisms, including runtime configuration, deployment, switch-over will be vital to be orchestrated across the infrastructure. 

Dynamic deployment is achieved by integrating continuous deployment technology on the edge of the network while coping with IoT's intrinsic heterogeneity~\cite{Yigitoglu2017,liyanage2016mepaas}. As IoT devices may be offline for quite some time, an orchestrator needs to gracefully cope with loss of connectivity~\cite{Rotsos2017} and increased likelihood of failing devices~\cite{Wen2017}.
 
Meeting the need of dynamic reconfiguration at the network level can increase network incidents and temporary malfunctions. A service orchestrator will also have to provide network diagnosis and root cause analysis during service disruptions~\cite{Rotsos2017}. In parallel, the orchestrator must support network resource scheduling that can adapt to near real-time service demands~\cite{Velasco2014}. \cite{Pradhan2017} focus on reliability of orchestration for IoT domains, proposing autonomous mechanisms that enable the ``analysis and management of:
\begin{inlinelist}
    \item the overall system goals describing the required applications, 
    \item the composition and requirements of applications, and 
    \item the constraints governing the deployment and (re)configuration of applications''.
\end{inlinelist}
 
Service oriented orchestrators for network functions have been proposed~\cite{Rotsos2016, Martini2016}. New techniques aim at integrating application information into orchestration decisions~\cite{Nadgowda2017}. Responding and adapting to specific application needs, while optimising resource usage can be an impossible mission and has been tried before with little success in practice.
 
IoT services can often be choreographed through workflow or task graphs to assemble different IoT applications~\cite{Peltz2003,chang2017mobile}. In some domains, the orchestration is supplied with a plethora of candidate devices with different geographical locations and attributes. In some cases, orchestration would typically be considered too computationally intensive, as it is extremely time-consuming to perform operations including pre-filtering, candidate selection, and combination calculation while considering all specified constraints and objectives. Static models and methods become viable when the application workload and parallel tasks are known at design time. In contrast, in the presence of variations and disturbances, orchestration methods typically rely on incremental scheduling at runtime (rather than straightforward complete recalculation by rerunning static methods) to decrease unnecessary computation and minimise schedule makespan~\cite{Wen2017}.

\cite{munozFrutos2009} propose a semantically enhanced mechanism to define quality-of-service for web services  (see Table~\ref{table:acc}). Similar techniques are likely to become more pervasive, but creating, adapting, and adhering to fixed ontologies has not proven highly effective. 

There are operational needs about the speed at which the orchestrator solver can process incoming monitoring information and return a fast and accurate enough decision. \cite{Amjad2017} focus on softening the orchestration decision making process to cope with scale. \cite{Pradhan2017} to formulate Satisfiability Modulo Theories (SMT) constraints that define desired system properties, enabling the use of SMT solvers to adaptively compute optimal system (re)configuration at runtime (see Table~\ref{table:acc}).
 
\subsection{Large(r)-scale and Fine(r) Grain }
As more devices are connected to edge networks and fog environments and the size of the unit of execution decreases (as described above) \cite{Jiang18}, it will be more difficult for cloud orchestrators to make a decision before the information they rely on becomes obsolete.

Service description must support aggregation/abstraction of resources in some ways to help with scalability (e.g. using hierarchical models), but the descriptions will have to be more abstract to cope with more devices and finer-grained execution units (with smaller state) --  see Table~\ref{table:acc}. \cite{Consel2017} offer a high level declarative language to describe implementation heterogeneous devices. Abstracting masses of IoT devices with heterogeneous capabilities remains a hard problem.

The scale that the fog/IoT impose on next generation orchestrators calls for mechanisms to describe the way data is handled, as the interplay between the swarm of devices executing the application and the data becomes more critical to achieve the required performance goals (see Table~\ref{table:acc}). \cite{Consel2017} suggested a design-driven approach that can be leveraged in two ways: first, design declarations are used by a compiler to generate a customised programming framework. These declarations can be supplemented with information to expose parallelism and allow efficient processing of large data sets.

Fog colonies/edge clouds may be distributed across a rather large area, interconnected through heterogeneous networks, while cloud resources are usually placed in centralised data centres. Discovering, selecting, and deploying devices can be built in 3 different ways: 1) hierarchical name system (like the Domain Name Service); 2) in an unstructured P2P flooding fashion (``ask your neighbour''); 3) hybrid (P2P at the edge), hierarchical there after (relying on nearby cloudlets~\cite{Satyanarayanan2009} and using central cloud services as a last resort).

Declarative model-based languages have been there for a long while and they are used by developers to express their resource needs and define preferred configurations in a more generic manner, rather than specifying the individual configuration of millions of devices~\cite{Goldsack2003, Bokharouss2008}. Mapping from these high-level configuration languages into finer-grained tiny units of execution is an open research challenge, but some solutions are already under way: \cite{Consel2017} defined DiaSpec, a declarative language to describe the functionality of an IoT device, abstracting over the specific hardware and  implementation. These declarations consist of source and action facets depending of the functionalities to be described. Each device of that type needs to conform to the interface and implement the sources and action operations. They also define a set of higher level constructs to work with large masses of sensors. 
 
Since resources are disaggregated, there is more flexibility and less cross-configuration interactions (e.g. a networking configuration affecting storage read/write throughput) but orchestrators need to become more robust. 
 
Configuration consistency also becomes an issue. An example would be setting up a high-speed channel between two functions executed in two different continents by orchestrating serverless environments as well as network control plane (say some traffic engineering is needed) and modifying VNFs along the data path (e.g. opening firewalls transiently). If the changes are not properly orchestrated, a sending function may send data through a data path that may filter or slow down that type of traffic. Synchronised clocks can be used to reduce the probability of having a violation of external consistency~\cite{Liskov1993}, but atomic clock synchronisation may be required for extremely latency sensitive orchestrations~\cite{corbett2013spanner}. 

\cite{Roca2017} define virtual fog functions (VFF) and several strategies to deal with mismatches between VFF and the capabilities of the underlying hardware in the IoT devices. This work suggests that a more interactive generation of orchestrators is needed, where the developer is in the loop at least at deployment/configuration time. 

Osmotic computing considers computational infrastructure as a chemical solution whose properties can change over time, the focus is on identifying the properties of what constitutes a solute and solvent, which is then operated on the principle of osmosis to manage and control services~\cite{Villari2016}  (see Table~\ref{table:acc}). Identifying how microservices can be migrated from edge resources to cloud-based resources (and vice versa), and what are characteristics influencing such migration, remains a challenge. The right formats and protocols for this to happen and cope with churn are not yet clear~\cite{Sharma2017}.

\subsection{Speed}
FaaS allows fine-grained, highly dynamic configuration. FaaS divides microservices in smaller software chunks that can be executed very quickly (price based on cpu/mem usage is a strong incentive to optimise function execution).

Smaller execution units that complete in seconds are a better fit for short lived resources, where failure is commonplace. This is a challenge for orchestrators, that need to decide where to execute a given FaaS function and reschedule (or take preemptive executions) to cope with failure. Slow, batch-style global optimisation is no longer an option. Instead, online-style techniques, with deadlines need to be explored in order to take advantage of the flexibility of functions. 

A main feature of serverless computing architectures is the ability/need to deploy new instances in the time scale of ms~\cite{Hendrickson2016}. This is also true for supporting flash events where millions of customers hit a website for a specific sales promotion, for instance. NFV functions need to be deployed/undeployed in sub-second time periods. 

Containers are the basic unit of deployment for serverless and many NFV functions. Container-based FaaS services tend to reuse the same container to execute multiple functions, even with this optimisation, serverless functions are significantly slower than containers at low request volumes~\cite{Hendrickson2016}. Some tricks to avoid the overhead of using persistent block stores to fetch data and configuration are possible. A scheduler aware that two different functions rely heavily on the same packages can make better placement decisions.

Session locality is an important factor (see Table~\ref{table:acc}): if a function invocation is part of a long-running session with open TCP connections, the orchestrator should run it on the machine where the TCP connections are maintained (avoiding traffic diversion by proxies). Mobility management however brings additional challenges to the orchestrator. 

Also, data locality will be important for running serverless functions pulling/pushing or scanning through massive state  (see Table~\ref{table:acc}). Orchestrators may require prediction capabilities to anticipate what data a particular function will read, making sure it is available to be function on time. 

All these problems with data and code locality exist at a data centre level, but they become more critical at the edge of the network. Advanced techniques to determine which edge nodes should be used to share the workload with and how much of the workload should be shared to each node are needed, heterogeneity and churn being the main deployment challenges~\cite{Jonathan2017}.

\subsection{Chaining Heterogeneous Functions and Storage}

IoT infrastructures are often modelled as a dynamic graph~\cite{Wen2017}. IoT configurations can be seen as a graph where the nodes represent the configuration and the edges the dependencies between task. Graphs are also used to describe virtual functions in NFV environments (see the NFV Management and Orchestration spec) and some serverless environments too (e.g. Oracle's Flow Fn serverless orchestration). This feels like a very intuitive approach, but some other serverless vendors orchestrate functions using pre-defined state machines (e.g.~\cite{Karl2016devops,StepFunctions2017}). 
 
In principle a similar approach can be used to declaratively describe the high-level features of function compositions in a fog environment, delegating lower level details to the orchestrator. The orchestrator would need to  map these high-level descriptions to vendor specific implementations (e.g. `key-value store' would map to different AWS or Google Cloud products). Locality and heterogeneity however bring additional challenges to the orchestrator.
  
The standard model of cloud data stores abstracts the physical location where information is stored. However, in a fog environment, together with the mentioned architectures that disaggregate storage from computation, information will be more fragmented than before, and the actual location where these (potentially small) data elements reside can be critical. Hence, location information needs to be included into the high-level descriptions used by orchestrators. 

Blending out together FaaS and NFV functions and their interactions with storage can be difficult (e.g. serverless does not support the same conventions that MANO NFV does). An orchestrator will need to not only have a compatible high-level description for all these elements, but also will need to have the right network access rights for interacting with each sub element of the system.

\subsection{Fine-grained Locality}

As the way to develop applications changes towards microservices, and FaaS, these concepts bring additional questions to how to discover existing functionality.  

The discovery of \textit{ad-hoc} services and available computing resources in the fog/edge needs to go beyond beyond predefined contracts and addresses. The probability of trying to contact a device that is no longer available is much higher at the edge, making device/service registries very ineffective. Mechanisms to initiate local p2p-style resource discovery at the edge have been suggested~\cite{Wen2017}, getting inspiration from the world of \textit{ad-hoc} networks.

Addressing end point mobility with session continuity is another solution to the problem of naming an churn~\cite{Yannuzzi2014}. The Locator Identifier Separation Protocol (LISP) allows an endpoint to switch between networks while keeping its Identifier (IP address) intact by maintaining the Routing Location (RLOC) of each Identifier in a mapping system, which is updated by its control plane. Also, Multi-Path TCP (MPTCP) defines TCP sub-flows at the transport layer based on the IP addresses of all the enabled interfaces on a device. Under mobility, whether the device changes its IP or switches radio technologies (e.g., WiFi to 4G), the new IP address is registered and a new sub-flow is opened. This strategy allows for seamless mobility of the device across networks and radio technologies  (see Table~\ref{table:acc}). 

The selection of the most appropriate function to be used is going to depend on where it needs to be executed. For instance, mobile services change physical locations and may require resources \textit{en route}. Thus, more expressive description mechanisms are needed to define these situations (e.g. hardware dependencies).

As for deployment and configuration, edge devices also need to be able to self-manage with little coordination from a central cloud location. \cite{Mohan2017} have recently developed a system based on edge communications with minimal cloud-driven coordination. The authors rely on edge (locally cached) content as clustering classifier, where a local leader coordinates communications for data retrieval and update. In a fog environment, a coordinator can be in the nearby fog layer, so as to cope with edge device churn.

As mentioned above, deploying lightweight monitoring modules that interact with the orchestrator, but do not overburden the edge devices, seems essential to reach a fair balance between synchronisation and network load.

\subsection{Multi -organisation/-tenant Orchestration}

Next generation clouds have to orchestrate resources from multiple administrative domains, this can be seen as an extension to the edge/fog and volunteer computing paradigms. While there are companies providing single domain facilities, the need to cross administrative orchestration has become much more pressing to take advantage of these latest trends. Cloud standards have failed to gain traction, but the need to find mechanisms for bridging the heterogeneity gap between platforms, and enabling data integration are more relevant than ever.

Most orchestration technologies working across administrative domains use a broker to orchestrate resources at different levels within a provider (e.g. the cloud and the edge network) and across providers (see~\cite{Bonafiglia2017} for a recent example). Broker models across providers and multi-stage schedulers and optimisers have been quite common in distributed computing and networking since at least 20 years ago~\cite{Nahrstedt1995}. 

Brokers have also been recently suggested as a viable model for cloud orchestration~\cite{Jrad2013,Samreen2016Daleel}. As the number of cloud vendors is limited, it is possible to build adapter and brokering layers that tried to homogenise access to different clouds. However, the hyper-heterogeneity and massive scale of edge/fog deployments makes this approach unfeasible. 

Network functions can be dynamically discovered, negotiated and elastically composed as services, application service providers may lease VNF chains with given communication capabilities from different ISPs and compose them to operate an end-to-end virtual service infrastructure to offer value-added application services to users (e.g., delay-optimised infrastructure for high-definition video applications~\cite{Latre2014}. 
 
One open question in most academic works is how to handle multi-tenancy and how to scale identity management services to a global scale~\cite{Bernabe2017} (see Table~\ref{table:acc}). Another often overlooked aspect is that at any point in time, some devices may belong to multiple organisations at once (not all users from the same organisation).

\begin{table*}[]
\centering
\caption{Mapping of how recent research contributions in the area of orchestration can contribute to the requirements identified above (in brackets). * means it applies to all requirements above.}
\label{table:acc}
\scriptsize
\begin{tabular}{c}
\toprule
\textbf{Recent Accomplishment} \\ \midrule
 Expressive declarative descriptions \\ \textbf{[heterogeneity, dynamism, churn, larger-scale, finer-grain]}~\cite{Goldsack2003,Banikazemi2013,Consel2017} \\ 
  State-machine based function orchestration definitions \\ \textbf{[dynamism, churn, larger-scale, finer-grain]} ~\cite{StepFunctions2017} \\ 
    Dataflow-based function composition \textbf{[*]} ~\cite{Jonas2017,Ravindra2017} \\ 
    Decoupling name from resource (LISP/ROC/MPTCP) \textbf{[*]} ~\cite{Jonas2017,Ravindra2017} \\ 
    Semantic quality of service \textbf{[dynamism, larger-scale]}~\cite{munozFrutos2009} \\ \midrule
    Spliting services into finer grained functions (``FaaSification'') \textbf{[finer-grain]} ~\cite{Dalla-Costa2017,Spillner2017} \\ 
    Data-aware config \textbf{[*]}~\cite{Dalla-Costa2017,Rehman2017} \\ 
    Trust-management, AAA, Channel-Protection \textbf{[Security]}~\cite{Bernabe2017,Stojmenovic2014, etsi-trust, Law2013}\\
    Serverless/FaaS isolation \textbf{[Security]}~\cite{sun2015}\\
    Security coordination (e.g. AAA, Trust) in Software Defined Storage \textbf{[Security]}~\cite{DARABSEH2017407}\\ \midrule
    Fluid (edge-fog-cloud) resource allocation/coordination \textbf{[*]}\\
    \cite{Villari2016,GilHerrera2016,Mijumbi2015,Munoz2015,Wen2017,Yannuzzi2014, Bonafiglia2017,Truong2015,Cirani2015,Rehman2017,Sehgal2015,Suciu2015}\\ 
    Working under different communication models (edge - p2p; fog - hierarchical; cloud - centralised) \\ \textbf{[dynamism, churn, larger-scale, multi-domain]}~\cite{Wen2017,Rehman2017}\\ 
    State (device and service) prediction \textbf{[dynamism, churn, larger-scale]}~\cite{Wen2017,Jonathan2017} \\ \midrule
    Softened goals in service/function composition/configuration \textbf{[larger-scale]}\\
    \cite{Goldsack2003,Huin2017,Amjad2017,Anderson2002,Wen2017} \\ 
    Fluid (edge-fog-cloud) resource allocation/coordination \textbf{[*]}\\
    \cite{Villari2016,GilHerrera2016,Mijumbi2015,Munoz2015,Wen2017,Yannuzzi2014, Bonafiglia2017,Truong2015,Cirani2015,Rehman2017,Sehgal2015,Suciu2015,Tuncer2015}\\ 
    Failure-tolerant orchestrator, cope with stagglers \textbf{[dynamism, churn, larger-scale, muti-domain]}~\cite{Brito2017}   \\ 
    Brokered (multi-domain) hierarchical orchestration~\cite{Jrad2013,Rotsos2016, Rostami2016, ohlen2016, John2017, Bonafiglia2017}\\ 
    Service support in orchestration decisions\textbf{[*]}~\cite{Nadgowda2017} \\ 
    Developer support in orchestration decisions(``device in the loop'') \textbf{[*]}~\cite{Roca2017,Consel2017} \\ 
    Self-protection, self-healing, self-repair, DoS protection \textbf{[Security]}~\cite{anastaciaconf1, santos2016selfnet, 7847329, bila2017}\\ 
    Universal identity management \textbf{[dynamism, churn, larger-scale, multi-tenancy, privacy]}~\cite{Bernabe2017}\\ \bottomrule												
\end{tabular}
\end{table*}

\subsection{Security and Privacy}

SDN-based IoT and Fog networks are vulnerable to the new-flow attacks, which can disable the SDN-based IoT by exhausting the switches or the controller. In this sense, \cite{7847329} authors present a smart security mechanism (SSM) to defend against New-Flow Attack in SDN-Based IoT differentiating new-flow attack from the normal flow burst by checking the hit rate of the flow entries.

Regarding security in SDStorage, in \cite {DARABSEH2017407}, a software defined based secure storage framework is proposed. Every storage control and security mechanisms are abstracted out from the hardware devices in the data plane and set inside the controller, enabling a centralised decision point based on security policies. Thus, when a host sends storage control packets and data traffic to another host in the network, the security controls such as authentication and filtering take place at the control layer instead of at the device level.

Regarding serverless and FaaS security, some initial works are starting to provide isolation at microservices and Serverless computing. Recently, Bila et al. \cite{bila2017} propose a policy-based improvement in the serverless architecture to guarantying and rebuilding vulnerable containers, that can be included as part of the security orchestration. Containers might have built-in vulnerabilities because of wrong configurations or just by the fact of including executable binaries with security flaws. In \cite{sun2015}, authors propose a Security-as-a-Service approach for microservices-based cloud applications, providing a flexible monitoring and policy enforcement infrastructure for network traffic to secure cloud applications. Unfortunately, there exist still few research and solutions aimed to cope with the emerging security issues in that field.

With regard to Edge and Fog computing, \cite{Stojmenovic2014} identified authentication at different levels of the gateways as the main security issue in fog computing. Multicast authentication~\cite{Law2013} or decoy information technology technique~\cite{Stolfo2012} have also been proposed to withstand malicious insiders by disguising information to prevent attackers from identifying customer's real sensitive data.

In \cite{Bernabe2017} authors explore the idea of privacy-preserving global identities that are universally valid for an entity. Such a feature is essential for handling authorisation in a large-scale distributed environment.

\cite{rlu2017} authors present a lightweight privacy-preserving data aggregation scheme, for fog computing-enhanced IoT that can aggregate hybrid IoT devices' data into one in some real IoT applications, so that user private data is concealed.

Recently, authors in \cite{gao} identified main attacks that can occur at the Edge-Fog and IoT, including, among others: DDoS, Routing attack, Sink node attack, Direction misleading attack, Black hole attack, Flooding attack, Sybil attack or Spoofing attack. To cope with those attacks the main countermeasures at network layer, focus, nowadays, on ensuring confidentiality, integrity and availability. To this aim, novel end-to-end encryption mechanism specially devised for IoT at different levels (e.g. 6LowPANs encryption, IPsec tunnels, DTLs), including Peer to Peer authentication and key negotiation management, can be orchestrated and configured on demand at the edge as security VNFs.

Trust-management in distributed scenarios such as inter-clouds have been addressed recently \cite{Bernabe2015}, where a semantic-web approach is followed to quantify dynamically trustworthiness and reputation among different clouds and services in order to establish reliable federations and communications among the parties. 

Malicious and curious adversaries (e.g. MEC data centres) can represent a privacy threat to the Edge/Fog users as they can gain some user-related information in the decentralised ecosystem.

In addition, new cybersecurity orchestration will need to provide self-protection, self-healing and self-repair capabilities through novel enablers and components \cite{santos2016selfnet} \cite{anastaciaconf1} at the Edge. To achieve those properties, services running in this environment need to work together with a lightweight (potentially distributed) watchdog that sends events to the orchestrator (e.g. compromised device triggers removal of keys and migration of data), to make reconfiguration decisions accordingly.

\subsection{Grouping Challenges}

The text in bold in Table~\ref{table:acc} shows a list of requirements that will be needed by next generation orchestrators. Most of these requirements have been tested in isolation. A more comprehensive orchestration approach joining all of them (and a few others that we highlight in the next section) would still be needed.

This conclusion arises from how orchestration has evolved in the last 20 years: orchestration challenges have evolved over time but mostly in separated areas that are now increasingly more unified due to the ''softwareisation" of IT. The orchestration needs can thus be classified in several waves:

\begin{enumerate}
    \item \textbf{$1^{st}$ wave}: software placement and communication in distributed (sometimes across domains) environments.
    \item \textbf{$2^{nd}$ wave}: same as $1^{st}$ wave but including edge and for resources in the IoT together with programmable networks and serverless functions.
    \item \textbf{$3^{rd}$ wave}: same as $2^{nd}$ wave but adding hardware programmability and disaggregation also at the edge and simplified data management (e.g. ~\cite{Lemos2015}).
    \item \textbf{$4^{th}$ wave}: same as $3^{rd}$ wave but with abstracting the underlying heterogeneity, complexity and dynamism of the IT infrastructure making it easier for human administrators and developers to use. 
\end{enumerate}

Taking this classification and Table~\ref{table:acc} into account, one could say that we are still in $2^{nd}$ wave orchestration technologies. 

The works in Table~\ref{table:conrevg} show how technologies are being integrated in pairs, with no efforts trying to cope with more than two at a time. This is a key characteristic of $2^{nd}$ wave orchestration technologies. Combining more than two orchestration comes with an exponential increase in complexity, which calls for new approaches towards comprehensive orchestration.

The next section presents some approaches to help us make the transition to $4^{th}$ wave orchestration faster and smoother.

\section{New and Revisited Orchestration Approaches}
\label{sec:appro}

\begin{table}[] 
\centering
\scriptsize
\caption{Pending challenges mapped to the requirements identified above (in brackets). * means it applies to all requirements above.}
\label{table:pend}
 \setlength\extrarowheight{-3pt}
\scriptsize
\begin{tabular}{cc}
\toprule
\textbf{Pending Element} & \textbf{Potential Solution} \\ \midrule
\thead{\scriptsize Abstract failure \textbf{[churn, dynamism]} \\ \scriptsize Data availability \textbf{[churn]} \\ \scriptsize Automated execution units description \textbf{[larger-scale, finer-grain]} \\ \scriptsize Hyper- heterogeneity \textbf{[*]} \\ \scriptsize Plain-English  searches \textbf{[*]}} & \thead{\scriptsize QoS enabled deployments, describe tolerable availability \\ \scriptsize Data-aware deployments~\cite{Lemos2015} \\ \scriptsize Automated  software splitting~\cite{Spillner2017} \\ \scriptsize Self-describing components \\ \scriptsize Information extraction and NLP} \\ \midrule

\thead{\scriptsize Device/service/function registries  \textbf{[*]} \\ \scriptsize Data directory  \textbf{[*]} \\ \scriptsize Matching resource requests and results \textbf{[dynamism,heterogeneity]} \\ \scriptsize Potentially O($N^2$) negotiation \\ \scriptsize \textbf{[churn, scale, multi-domain]}~\cite{Kelaskar2002} \\ \scriptsize Registry scalability \\ \scriptsize discovery/sharing functionally similar functions} & \thead{ -- \\ -- \\ \scriptsize Ontology-based searches~\cite{Sim2012} \\ -- \\ \scriptsize Decentralised (P2P) registries \\ \scriptsize creating libraries and packages of  functions} \\ \midrule
 
\thead{\scriptsize Slow selection \textbf{[dynamism, speed, larger-scale]} \\ \scriptsize Universal naming beyond LISP/ROC/MPTCP \textbf{[churn, multi-domain]} \\ \scriptsize Redundancy and ``high availability'' \textbf{[churn, dynamism]} } & \thead{\scriptsize Improve selection based on previous runs ~\cite{Bankole2013} \\ \scriptsize Logical resource names~\cite{Melander2011} \\ \scriptsize -- } \\ \midrule

\thead{\scriptsize Edge-fog-cloud coordination  \textbf{[*]} \\ \scriptsize  Orchestration always catching up \textbf{[churn, dynamism]} \\ \scriptsize Automated adaptation \textbf{[heterogeneity]} \\ \scriptsize Isolation in FaaS/Serverless  \textbf{[security]}
\\ \scriptsize Cloud/Edge/Fog Security Coordination through \\ \scriptsize NFV/SDN (Trust, AAA, ChannelProtection, key-management) \textbf{[security]}}
& \thead{\scriptsize Emergent behaviours, Asymptotic configurations~\cite{Goldsack2003,Huin2017,Amjad2017,Anderson2002} \\ \scriptsize Unsupervised learning of configuration options \\ \scriptsize ---\\  \scriptsize ---\\} \\ \midrule

\thead{\scriptsize Slow resource provisioning \\ \scriptsize  \textbf{[large-scale, finer-grain]}  \\ \scriptsize Deployment obsolescence  \\ \scriptsize \textbf{[dynamism, churn, larger-scale, finer-grain]} \\ \scriptsize Stateful workflows \\ \scriptsize \textbf{[dynamism, locality]}\\ \scriptsize Across provider federation \\ \scriptsize \textbf{[multi-domain, multi-org]} \\ \scriptsize Accessing byte addressable memory beyond data centres \\ \scriptsize \textbf{[*]}} & \thead{\scriptsize Predictive resource estimation~\cite{Bankole2013,Arabnejad2017} \\ \scriptsize Predictive scheduling~\cite{Berral2010} \\ \scriptsize Delegation, asymptotic deployment~\cite{Pollock1998, Anderson2002} \\ \scriptsize Data access prediction \\ \scriptsize Workflow delegation/handover \\ \scriptsize --} \\ \midrule

\thead{\scriptsize Data lifecycle management  \textbf{[*]} \\ \scriptsize Global workflow \textbf{[multi-domain, -org, larger-scale]} \\ \scriptsize Balance synchronisation and network load \\ \scriptsize \textbf{[dynamism, speed, churn]} \\ \scriptsize Secure orchestration  \textbf{[*]} \\ \scriptsize Automated control loop-based monitoring/re-config~\cite{Abid2017} \\ \scriptsize Limited programming models \\ \scriptsize \textbf{[dynamism, data lifecycle, churn]}~\cite{Latronico2015,Jonas2017,Consel2017} \\ \scriptsize Ability to debug and explain \textbf{[*]} \\ \scriptsize Autonomic security reconfiguration and orchestration \\ \scriptsize in SDN-NFV-enabled Fog and IoT \textbf{[security]} \\ \scriptsize Monitoring, accounting in FaaS/Serverless~\cite{Baldini2017} \\ \scriptsize \textbf{[security]}} & \thead{\scriptsize Data-aware orchestrator \\ \scriptsize Delegation \\ \scriptsize Statistical-monitoring - \\ \scriptsize operate on aggregated monitoring information \\ \scriptsize Constant influx of security/privacy watchdog \\ \scriptsize Automated loop constraint generation \\ \scriptsize HEBs~\cite{Roca2016} \\ \scriptsize Abstract description and heavy delegation \\ \scriptsize -- \\ \scriptsize IA-driven contextual monitoring/reaction \\ \scriptsize --}  \\  \bottomrule
\end{tabular}
\end{table}

\subsection{Learning to Orchestrate}
 
Machine learning (ML) techniques are starting to be applied to different aspects of the orchestration process in the cloud, such as data centre scheduling~\cite{Berral2010,Gao2014,Marco2017spark}, IaaS instance selection~\cite{Samreen2016Daleel}, optimising resource scalability~\cite{Bodik2009,Bankole2013,Arabnejad2017}, network flow classification~\cite{Zander2005,usama2017unsup}, network performance prediction~\cite{Stadler2017}, and software defect classification~\cite{Nam2015SWDefectClass}. When fed with the enormous amount of logs kept by data centre and network operators, ML models are able to select the best configuration and location of resources.

These could be applied to making better and faster resource selection and configuration in edge/fog/IoT environments, but the amount and quality of data required and the need to pull these data out of many different organisations make it less workable, at least for now. Additionally, technologies for combining ML models trained for different domains (NFV in a telco with SDN in a data centre and SDN in a backbone network, for instance) into a single workable solution need to be explored.
 
The combination of models does not need to be in a hierarchical fashion~\cite{Rotsos2016, Rotsos2017}, however. For instance, in the case of a local ML model trained to optimise optical interconnects in the transport network and getting requests from a peer data centre model to open up connections with minimum latency for a set of VMs hosting a bulk data transfer or a WAN acceleration VNF. A transport network-associated neural network and the data centre-associated neural network can automatically negotiate the best setup for that VNF based on prior instances of setting up that (or a similar) connection. These trained models can handle multiple objectives, like optimisation~\cite{Huin2017} and resiliency. Feature-Weighted Linear Stacking or buckets of models are commonly used techniques to combine models. We envision these neural net models can learn to talk to each other without specifying the details of the communication protocol (e.g. unsupervised learning of configuration protocols in Table~\ref{table:pend}).
 
Service discovery is more difficult to solve using ML techniques alone: services would need to be registered in some form of discovery service and tagged so that they can be found. This labelling needs to be done by highly qualified individuals, which makes the process tedious and not scalable. 
 
Device churn, on the other hand, would require highly generalisable models, trained under an incredible variety of circumstances in order to minimise runtime overfitting-derived errors resulting from training with a very specific snapshot of the system.

While standards are needed, forcing humans to use these often results in no-standard usage or the development of yet another new ``standard''. Higher level languages are needed to solve this standard proliferation problem.
 
Declarative model-based languages focus on configuration of resources and services~\cite{Goldsack2003,Bokharouss2008,Banikazemi2013,Consel2017}, but they are less useful when it comes to service discovery and composition. Developers rely on the usage of web search engines to find compatible services. Information extraction, natural language processing, data and mining and similar technologies will help in this regard, as indicated in Table~\ref{table:pend}. 

\subsection{P2P/Agent-based Orchestration}

P2P systems have traditionally excelled at delivering robust applications on vast numbers of edge devices with high volatility across multiple domains, but also within a single data centre for specific services (e.g. HPE Smart SAN\footnote{\scriptsize\url{http://h20195.www2.hpe.com/V2/getpdf.aspx/a00001440enw.pdf}}). 
P2P orchestration means independent agents that are capable of making autonomous decisions about a set of resources they control. These decisions are not necessarily prescribed by a set of immutable rules, but the agents adapt their strategies based on the state of the resources and the value of the applications they try to run on them. 
  
Descriptions tend to be based on pre-established ontologies~\cite{Sim2012}. Thus, it is very complicated to make agent-based systems negotiate about a new type of request or resource they have never seen before (not in the ontology), requiring constant updates and maintenance. Hence, coping with hyper-heterogeneity still seems like a hard mission (Table~\ref{table:pend}).
 
In the case of a fog coalition where individual devices all negotiate how to orchestrate a running application, agents negotiating in pairs may take long (potential poor scalability, O($N^2$) for a full mesh -- see Table~\ref{table:pend}) so some structure needs to be imposed to prevent long negotiation rounds. Moreover, high device churn would require negotiations to re-start. Also, discovering peers for negotiation can be complex under different domains~\cite{Kelaskar2002}.
 
A key goal for any orchestrator is to capitalise on low-level interfaces and synthesise new service-oriented abstractions that minimise human interaction and provision service in the order of minutes or seconds~\cite{Rotsos2016}. Taming service description and composition so that developers do not trade standardisation for convenience will drive a few research works in forthcoming years. 

P2P orchestrators have also been tried as an alternative to centralised brokered orchestrators for quite a long time~\cite{Chafle2004}, but they have found very limited success in practice. For instance, Netflix have decided not to use P2P task choreography in their Conductor microservices orchestration engine. P2P systems tend to create implicit contracts that result in poorly documented tight coupling around input/output, SLAs, etc, making it harder to adapt to changing needs. Controlling the deployment/management of a myriad of individual controlling agents and creating a hierarchy for debugging/delegation/escalation is a complicated task (Table~\ref{table:pend}). 
 
There is a wealth of knowledge about P2P/agent based systems and security, but some aspects introduced by the amalgamation of NFV/SDN/disaggregated data centres/FaaS are not well explored. Storing data from nearby devices may expose peers to legal liabilities that may hinder further developments.

\subsection{Eventually Consistent/Probabilistic Orchestration}
 
To ensure near-real-time intervention during IoT application development, one approach is to use correction mechanisms that could be iteratively applied even when suboptimal solutions are deployed initially. In this setting, the good old asymptotic and declarative management techniques~\cite{Pollock1998, Anderson2002} may likely be applicable to manage these highly complex scenarios. In the same vein, differential consistency techniques, where devices get serializable consistency only in their neighbourhood (vs eventual consistency for further devices) have been suggested for distributed data stores~\cite{Mayer2017}. Similarly, \cite{Abid2017} suggest the use of several concurrent control loops that are automatically generated from a simple description language, as a mechanism to achieve eventual consistency between a desired state of the resources and their actual state (Table~\ref{table:pend}). Manual constraint generation no longer seems feasible in the light of current multi-domain, hyper-heterogeneous, IoT scale trends. 

Massive scales, time uncertainty, resource dynamism, and delayed monitoring can all be coped with by applying asymptotic management techniques, as long as the application tolerates delays and performance does not degrade (or cost does not spike) quickly. Most current orchestration frameworks do not easily tolerate partial failure or undetermined delays to make resources gradually available and, thus, progressively take the system closer to the desired state (Table~\ref{table:pend}). 

\cite{Korupolu2016} provide a theoretical framework for the allocation of batch and service jobs in a set of constrained resources where some resources can be attacked or fail. Achieving a target reliability level can simply be a matter of placing extra replicas in different failure domains~\cite{Sedaghat2016}. Defining failure domains at the edge, especially with end-user devices, can be difficult and requires proper observability and late characterisation of the failure modes of the devices. 
Also, resource definition languages need to be made in terms of tolerable availability and needed capacity, so that the orchestrator can factor these in. Orchestrators would also benefit from some historical knowledge to apply correction factors depending on previously seen churn and failure rates.
 
\subsection{Hierarchical Delegation}
The presence of a common data model (semantically rich enough for expressing the required goals) and a common mechanism for labelling the entities in the model (so that information can be fed backwards once a delegated operation has been materialised) enables this information exchange. Delegation approaches rely on declarative languages used by developers to express their resource needs~\cite{Bokharouss2008}. 
 
Once the user specifies a set of elements to be deployed and how they are connected, the infrastructure labels each element with a unique name (Table~\ref{table:pend}). There is a degree of information that remains unknown for the user (e.g. underlying infrastructure details or topology of the virtual infrastructure, i.e. administrative domains and contract terms with each of these ones).
 
Some works use delegated workloads that are then analysed and scored before their allocation~\cite{Verma2015}. In spite of Google's workload and server heterogeneity, these are not comparable to the hyper-heterogeneity scenarios we described above. Also, dealing with orchestration of containers in a single domain helps make pragmatic decisions that work at scale in a well-confined administrative/security domain. 
 
In a NFV/SDN/Serverless/Edge/Fog/IoT scenario descriptions are refined, transformed, and split (e.g. across several domains).  It is possible that the configuration details for the edge nodes cannot be completed without information from the neighbouring domain. For example, they may need to exchange ports, IP address or tags depending on the nature of the connection. They may also need to agree which edge nodes to use. It is likely that this is information that they will not be willing to share with anyone other than their neighbouring domain. This implies the information will be obtained by interaction between providers and needs to be referenced differently as the model gets refined~\cite{Melander2011,Vilalta2016}.
 
Model refinement and splitting has a direct implication over the way things are referenced in the data model. For instance, when a user specifies she desires a VM in the UK and another one in the US, she is (likely) indirectly generating a split of her request across multiple administrative domains. When her request is split and refined, the VM in the UK is referring to a VM in the US whose final name (e.g. its host name) would only be known after deployment.
  
Hierarchical delegation models work well across administrative domains and their divide and conquer approach tends to enable larger scalability. Service discovery happens within a single administrative domain, where classic registry and search approaches have proven to work. However, implementing these systems in the light of hyper-heterogeneity is very difficult and expensive. They also fall short when it comes to coping with fluid dynamic organisations and high resource churn (Table~\ref{table:pend}).

\subsection{No Orchestration}
The best of orchestration might be having to do no orchestration at all. Large scale production systems call for very simple orchestration techniques where developers and operators can rapidly debug things gone badly (e.g. most cloud schedulers use simple round robin for resource allocation). Even if the right abstractions are provided, building resource requests in high-level declarative languages can prove to be tedious and error prone for most developers. Several approaches try to bring the orchestration problem closer to developers and expose interfaces that let programmers specify behaviour, while concurrency and access control are individually dealt with by different devices (see Table~\ref{table:pend}).
 
\cite{Zhang2015} propose a framework where each device publishes a global log (time series of data and events) that is readily available for actuators to use. The framework is, however, too high level and does not really define how it would cope with trillions of time series. 
 
Swarmlets are presented as an elegant way to use the actor model to wrap access to sensor devices~\cite{Latronico2015}. Thus, developers would simply instantiate accessors on devices, reducing the orchestration needs. Either resources themselves or developers would have to control access, especially when modifying configurations. Indeed, it seems, Swarmlets add one level of indirection but similar questions as to how to publish, register and search for accessors; accessor security or lifecycle remain. 
 
Fn Flow and PyWren~\cite{Jonas2017} agree that the best approach to orchestrating FaaS is using familiar programming models (Java 8 lambdas and BSP or M/R, respectively). The description of the code functions would be done in the development environment and selection will come as using any library with dot autocomplete, leaving configuration, monitoring, and deployment to the underlying middleware. There are questions as how these libraries of functions could get organised and be made accessible for thousands of remote development environments without damaging developers' experience (see Table~\ref{table:pend}). There are also question marks on how this approach could cope with hyper-heterogeneity and edge deployments.

Beyond devices (or infrastructure) themselves, there is the problem of discovering many small fine-grained functions that could be (re)-used by multiple applications. There are no well-defined patterns for discovery across FaaS functions. While some of this is by no means FaaS specific the problem is exacerbated by the granular nature of FaaS functions and the lack of application / versioning definition. In this sense, building on classic software engineering practices (creating libraries and packages of functions) may be the way ahead (see Table~\ref{table:pend}).

Fog Dataflow programming frameworks have also recently appeared that support developers in dealing with scalability, heterogeneity, and mobility~\cite{Giang2015}. They do not support the levels of device autonomy, churn and hyper-heterogeneity (while keeping low levels of human intervention) we have described above.
 
Moreover, all of these approaches are very interesting for small homogeneous deployments confined to a single domain. The levels of complexity we will see in hyper-heterogeneous large scale fluid distributed organisations seem too complex for any single unassisted developer to cope with.

\subsection{Security Orchestration} 

New context-aware holistic security orchestrators are needed to allow interfacing with NFV managers, SDN controllers and Edge-Fog controllers, thereby providing security chaining, as well as dynamic reconfiguration and adaptation of the virtual security appliances at the edge, in case of deviation from the expected behaviour. 

In this sense, new security orchestration approaches are appearing recently. Open Security Controller (OSC)\cite{osc} is an open source project that tries to provide consistent security across a multi cloud environments. It aims to automate the deployment of virtualised network security functions to protect east-west traffic inside the data centre. It orchestrates the deployment of virtual network security policies, applying the correct policy to the appropriate workload.

Security Orchestrator \cite{Jaeger} proposes a design of a Security Orchestrator in the context of the ETSI NFV Reference Architecture, defining the interfaces required  to  interact  with  the  existing  MANO entities. The Security orchestrator is placed outside the architecture to achieve a holistic end-to-end security view  in  case  of  a  hybrid  network.  

In order to mitigate cyber-threats, latest research efforts focus on providing dynamic, intelligent and context-aware security orchestration in Fog/Edge and IoT by relying on NFV/SDN-enabled networks. This approach allows chaining and enforcing policy-based security mechanisms while providing run-time reconfiguration and adaptation of security enablers, and therefore, endowing the ecosystem with intelligent and dynamic behavior. In this sense, the H2020 EU project Anastacia \cite{anastaciaconf1} is also devising a security orchestrator to take autonomous decisions in MEC, Cloud and IoT scenarios, through the use of networking technologies such as SDN-NFV and intelligent and dynamic security policy enforcement and monitoring methodologies. In the Anastacia project, different virtual security appliances such as vFirewall, vIDS, vAAA, vSwitch/Router, vHoneynet, vVPN are orchestrated dynamically at the edge of the network.

In order to achieve a context-aware autonomic security orchestration in SDN/NFV-enabled Fog and IoT, we envisage the proliferation of cyber-situational awareness frameworks in which the security orchestration can dynamically be adapted according to the context obtained from agents and sentinels, mitigating and countering cyber security threats at the edge, by deploying and orchestrating Virtual Security Functions and services even in constrained Fog and IoT devices. Such awareness framework could be endowed with monitoring and reaction tools as well as innovative algorithms and techniques based on machine learning, for threat analysis, data fusion and correlation from different sources, and big data analysis. It would allow to increase the overall security, including self-repair, self-healing and self-protection capabilities, not only at the core, but also at the edge of the network.

\subsection{Hierarchical Emergent Behaviours}
A recent paper proposed Hierarchical Emergent Behaviours (HEB), an architecture that builds on established concepts of emergent behaviours and hierarchical decomposition and organisation. HEB's local rules induce emergent behaviours, i.e., useful behaviours not explicitly programmed~\cite{Roca2016} (see Table~\ref{table:pend}). This certainly is a promising approach, however it hinges heavily on the availability of an accurate and detailed model of all the resources available to an orchestrator, including all elements of the underlying infrastructure of available functions, resources, and deployment locations. Mechanisms to acquire such a model are not yet available in the literature, and is something that we hope to be closer to by addressing the challenges discussed thus far.
 
Emergent behaviours eliminate the need of a central orchestrator that would have to deal with a very large number of ``things''. They still require the use of high-level languages and associated tools (including ontologies) to describe the emergent behaviours~\cite{1413118,Blair2015holons,Blair2016,7774391,Oquendo2017}.
 
Due to the large number of variables and situations, designing an explicit programmed system that takes into account all the scenarios in advance is a formidable task. With an HEB IoT approach and if the proper set of rules is defined, the ``things'' are able to dynamically adapt to the environment without the need to explicitly program them. However, they also extend the ``attack surface'' that can be exploited. An attacker that gains access could modify the rules, either directly or through modification of the hyper-parameters, for nefarious purposes.

\section{Conclusions}
\label{sec:fin}
Recent technological developments have been too quick for orchestration techniques to catch up. Most current orchestrators would fit in the $2^{nd}$ wave orchestration classification above, as they either do not cope with hardware programmability and disaggregation beyond a data centre. 

Moreover, the amalgamation of many virtualised and disaggregated technologies is making end-to-end orchestration difficult to do at scale. This situation is getting worse with the advent of the IoT, where hyper-heterogeneity can make it nearly impossible for a single team to master all the knowledge needed across the stack. Scale is not the only problem, churn and dynamism makes it very hard to discover resources or plan how to synchronise as many of these devices connect to the network intermittently only and/or belong to different administrative domains.

Pushing current orchestrators to the next wave of maturity calls for further integration across (hardware) and ''along" the technological spectrum these technologies cover. Recent works have made the challenges we identified less intimidating, offering promising results to tame data aware deployments, resource planning at scale (e.g. asymptotic plans), coping with churn (HEBs, logical naming, predictive scheduling) or resource searches (dynamic ontologies). 

There is, however, lots of pending work to do if we aim for integrated solutions that can deliver a unified approach to orchestrate across technologies and administrative domains. Achieving  $3^{rd}$ wave-level orchestrators requires better automation and complexity abstraction techniques and systems that can make automatic, but adaptive, decisions based on as few human inputs as possible. To a human developer or administrator, this mix of technologies needs to look as if it was a simple local deployment using a very declarative form, vs. very prescriptive specifications. Thus, artificial intelligence, NLP, HEBs, or ''No orchestration" techniques can help smooth the interface between humans and this massive complexity, which are essential for $4^{th}$ wave-level orchestrators to become a reality.

\section*{Acknowledgement}
The authors thank Gordon S Blair for his insightful comments on prior versions of this manuscript.

\bibliographystyle{plainnat}
\bibliography{orch}

\end{document}